\documentclass[3p,times,12pt]{elsarticle}
\usepackage{amssymb}
\usepackage[figuresright]{rotating}
\usepackage{color} 
\usepackage{url}

\newcommand{\be}{\begin{equation}}
\newcommand{\ee}{\end{equation}}
\begin{document}
\begin{frontmatter}
\title{Helical solitons in vector modified Korteweg--de Vries equations}

\author{Dmitry E. Pelinovsky$^{a,c)}$, Yury A. Stepanyants$^{b, c)}$\footnote{Corresponding author,
		e-mail: Yury.Stepanyants@usq.edu.au}}%

\address{$^{a)}$ Department of Mathematics and Statistics, McMaster
	University, Hamilton, Ontario, Canada L8S 4K1 \\%
	$^{b)}$ Faculty of Health, Engineering and Sciences,\\ University
	of Southern Queensland, Toowoomba, QLD, 4350, Australia \\%
	$^{c)}$ Department of Applied Mathematics, Nizhny Novgorod State Technical University, Nizhny Novgorod, 603950, Russia}

\begin{abstract}
	We study existence of helical solitons in the vector modified Korteweg--de Vries (mKdV) equations, one of which is integrable, whereas another one is non-integrable. The latter one describes nonlinear waves in various physical systems, including plasma and chains of particles connected by elastic springs. By using the dynamical system methods such as the blow-up near singular points and the construction of invariant manifolds, we construct helical solitons by the efficient shooting method. The helical solitons arise as the result of co-dimension one bifurcation and exist along a curve in the velocity-frequency
	parameter plane. Examples of helical solitons are constructed numerically for the non-integrable equation and compared with
	exact solutions in the integrable vector mKdV equation. The stability of helical solitons with respect to small perturbations is confirmed by direct numerical simulations.\\
\end{abstract}


\end{frontmatter}

\newpage%
\section{Introduction}
\label{Sect1}

There are several different types of solitons in physical systems,
including classical Korteweg--de Vries solitary waves and their
generalisations with the exponential, algebraic, and oscillatory
asymptotics, as well as kinks, envelope solitons, two-dimensional lumps,
topological solitons, breathers, etc. (see, e.g., \cite{Lamb1980, Ablowitz1981}). Less known are
helical solitons which appear in vector models of nonlinear
equations.

One of the first examples of helical solitons was
reported in Ref.~\cite{Gorshkov1974} for circularly polarised
waves in solid-state plasma described by a rather specific nonlinear
wave equation. Other examples of vector equations describing
plasma waves were considered in late 1970s by several authors
\cite{Kuehl1977, Shukla1977, Dysthe1978} who derived a vector
modified Korteweg--de Vries (mKdV) equation for the description of
small-amplitude long waves. This equation in the dimensionless form is:
\begin{equation}
	\label{Eq01} %
	{\bf u}_t + (|{\bf u}|^2 {\bf u})_x + {\bf u}_{xxx} = 0,
\end{equation}
where ${\bf u} = (u_1, u_2)$ is a two-component vector with the norm $| {\bf u} | = \sqrt{u_1^2 + u_2^2}$.
In some cases a similar equation can be derived with the negative dispersion coefficient; we do not consider such cases here.

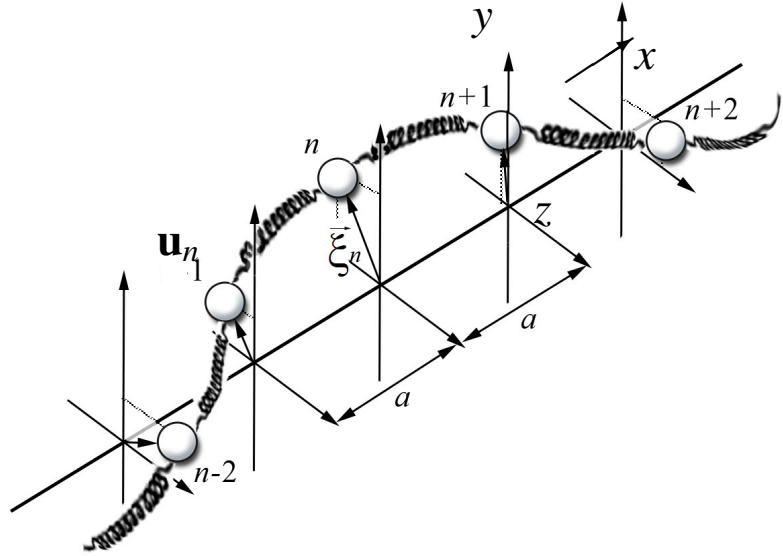
\begin{figure}[h]
	{\centerline{\includegraphics [scale=0.25]{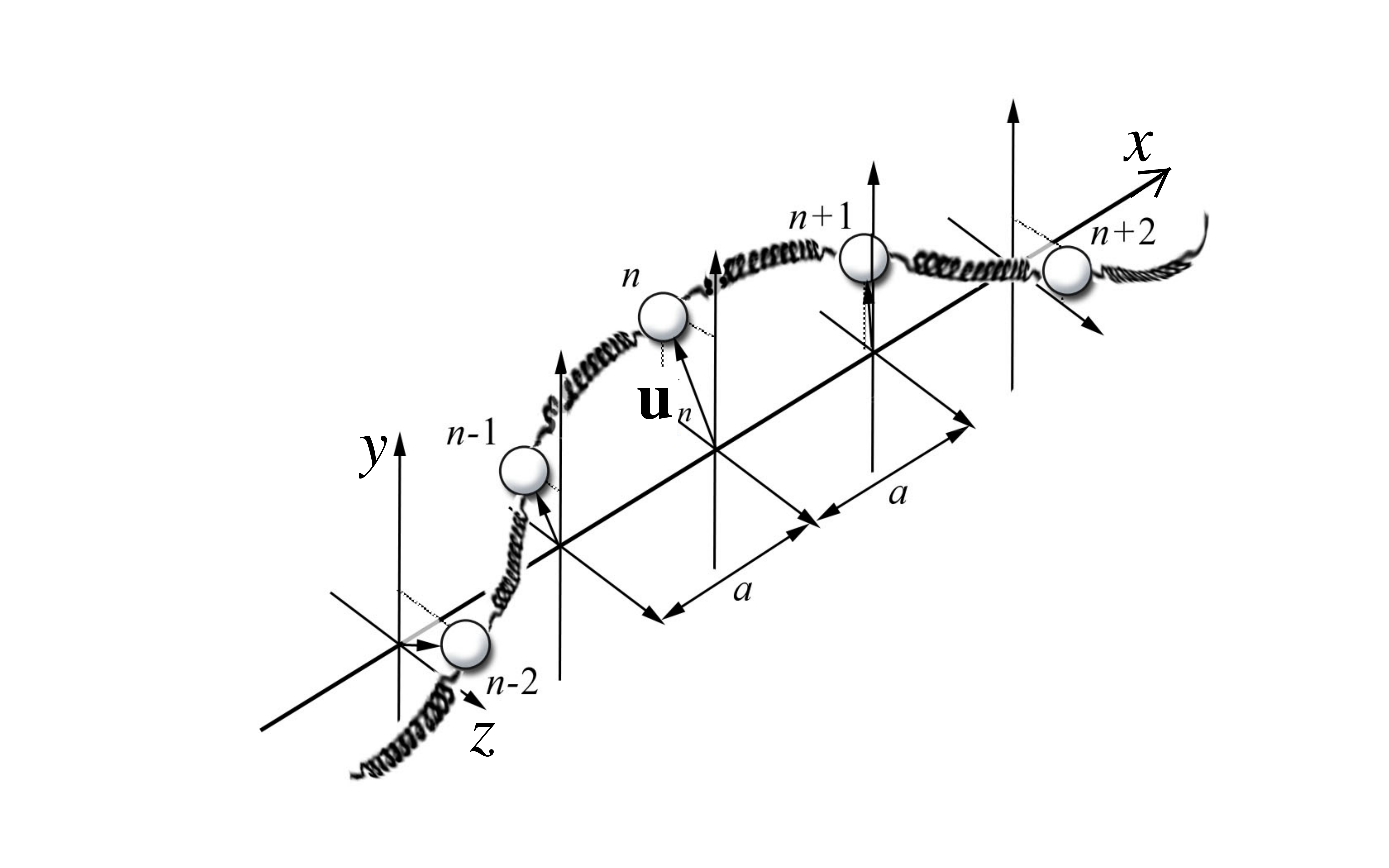}}} %
	\begin{picture}(300,6)%
	{\thicklines %
		\put(159,145){\Large \colorbox{white}{\thicklines{\phantom{\line(0,1){20}}}}}%
		\put(162,145){\Large \colorbox{white}{\thicklines{\phantom{\line(0,1){20}}}}}%
		\put(165,157){\Large \colorbox{white}{\thicklines{\phantom{\line(0,1){4}}}}}%
		\put(335,230){\Large $x$}%
		\put(275,245){\Large $y$}%
		\put(298,172){\Large $z$}%
		\put(157,160){\Large ${\bf{u}}_n$}%
		\put(310,225){\vector(3,2){25}}}%
	\end{picture}
	\vspace{-1.5cm}%
	\caption{A schematic example of a transverse flexural perturbation on the chain of particles linked by elastic springs.} %
	\label{f01}
\end{figure}

Equation (\ref{Eq01}) was also derived for the
description of transverse perturbations in a chain of interacting
particles \cite{GorbOstr1983, Nikitenkova2015}, nonlinear waves in micropolar media \cite{Erbay1989}, in generalised elastic solids \cite{Erbay1998}, and deformed
hyperelastic dispersive solids \cite{DestrSacc2008}. Figure
\ref{f01} illustrates transverse flexural perturbations travelling along a
chain of particles connected by springs. The particle displacement
in the plane perpendicular to the axis of propagation (the $x$-axis) has
two components, $y$ and $z$, and can be presented by a two-component vector ${\bf u} = (u_1, u_2)$.

The vector mKdV equation (\ref{Eq01}) is non-integrable in contrast to its integrable counterpart:
\begin{equation}
	\label{Eq02} %
	{\bf u}_t + |{\bf u}|^2{\bf u}_x + {\bf u}_{xxx} = 0.
\end{equation}
Both these equations, Eq.~(\ref{Eq01}) and Eq.~(\ref{Eq02}), can be presented
in the scalar form for the complex variable $u = u_1 + i u_2$.
In the case of integrable vector mKdV equation, the scalar complex variable $u$
satisfies the complex mKdV equation, solutions of which can be constructed by the inverse scattering
method \cite{Karney1979}. Although the integrable vector mKdV equation (\ref{Eq02}) did not find any physical application,
the model can be considered as an asymptotic analog of Eq.~(\ref{Eq01}) for certain perturbations.

\begin{figure}[ht]
	{\centerline{\includegraphics [scale=0.35]{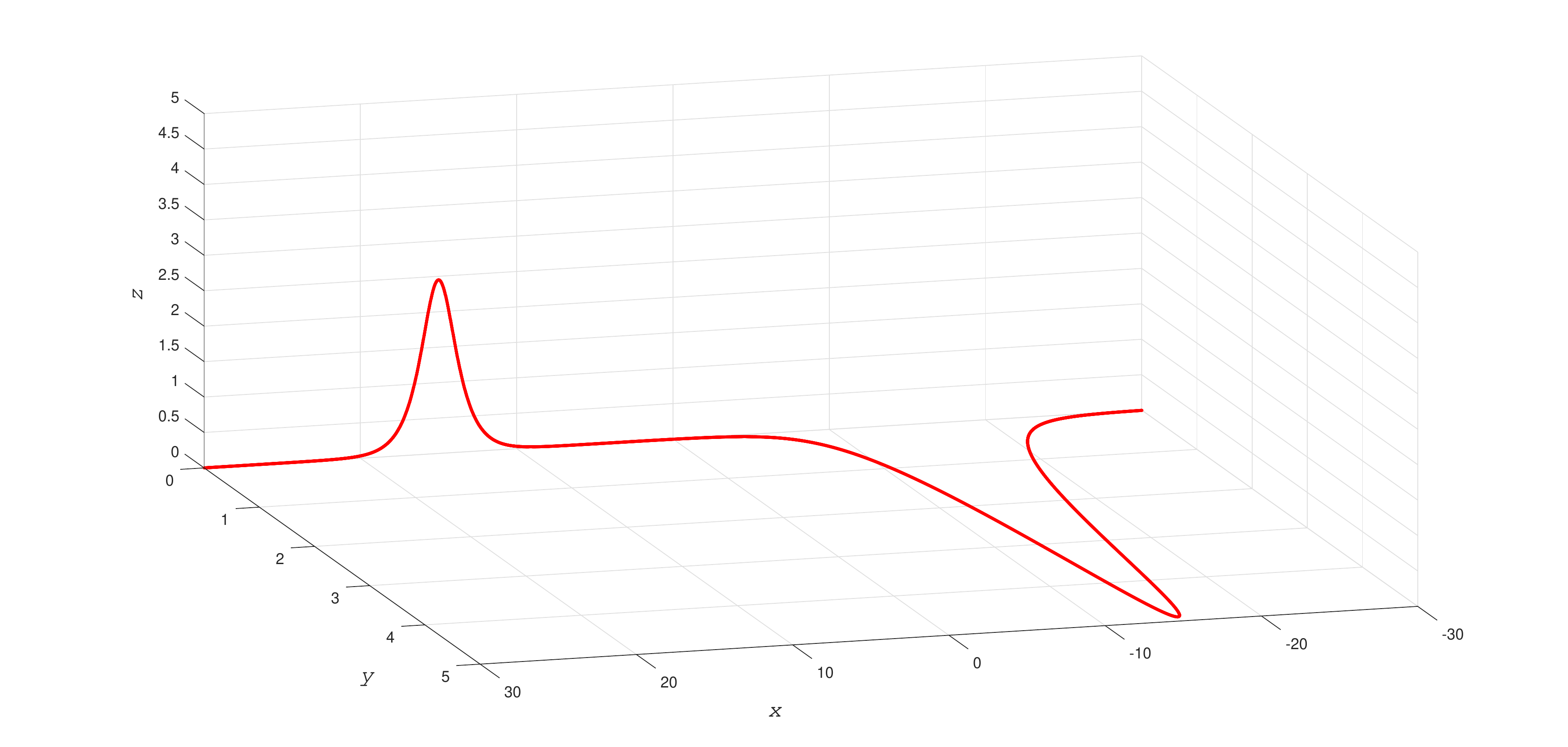}}}
	\vspace{-0.5cm}%
	\caption{Two planar solitons in the non-integrable vector mKdV equation (\ref{Eq01}) polarized
		in the perpendicular directions at the angles $\Theta_1 = 0$ and $\Theta_2 = \pi/2$.}%
	\label{f06}
\end{figure}

The traveling solitons of the vector mKdV equation (\ref{Eq01}) are solutions of the form:
\begin{equation}
	\label{trav-sol}
	{\bf u}(x,t) = {\bf U}(x-Vt)
\end{equation}
with ${\bf U}$ vanishing at the infinity and $V$ being the soliton speed.
When (\ref{trav-sol}) is substituted to Eq.~(\ref{Eq01}),
the resulting ODE can be integrated once with the zero constant of integration to the form:
\begin{equation}
	\label{TravSol} %
	{\bf U}_{xx} + |{\bf U}|^2 {\bf U} - V {\bf U}  = 0.
\end{equation}
This equation is rotationally invariant in the $(y,z)$ plane. Therefore, the solutions are polarized in a plane
with the displacement vector ${\bf U}$ given by
\begin{equation}
	\label{trav-sol-polar}
	{\bf U}(x-Vt) = r(x-Vt) [\cos\Theta,\sin\Theta],
\end{equation}
where $\Theta$ is a fixed polarization angle. For the solutions in the form (\ref{trav-sol-polar})
the vector equation (\ref{TravSol}) reduces to the scalar equation for $r$.
It is easy to show that the scalar equation for $r$ has the exact soliton solution for any $V > 0$.
Figure \ref{f06} illustrates two planar solitons initially polarized in the perpendicular planes.

Interaction between the traveling solitons with different polarizations
was studied numerically for Eq.~(\ref{Eq01}) in Ref. \cite{Nikitenkova2015}. The interaction was shown to be
inelastic, in general, except the trivial case when both solitons lie in the same plane;
in such case Eq.~(\ref{Eq01}) reduces to the scalar mKdV equation.

\begin{figure}[h!]
	{\centerline{\includegraphics [scale=0.35]{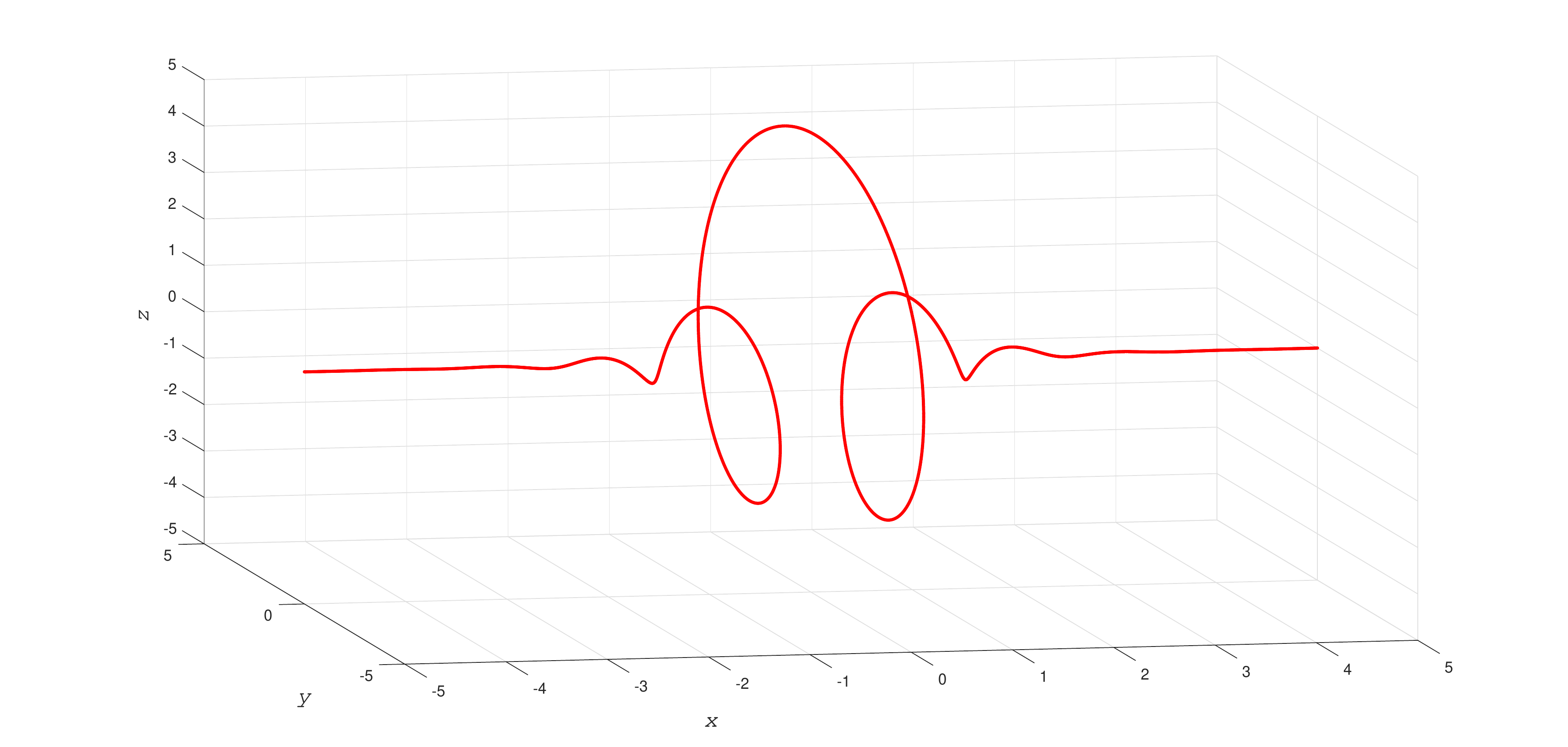}}}
	\caption{The helical soliton (\ref{helic-sol}) in the integrable vector mKdV equation (\ref{Eq02})
		with $V = -104$ and $\omega = 480$.}%
	\label{f02}
\end{figure}

The helical solitons of the vector mKdV equation (\ref{Eq01}) are solutions with the variable phases $\Theta$:
\begin{equation}
	\label{helic-sol}
	{\bf u}(x,t) = r(x-Vt) \left[\cos{\Theta(x, t)}, \sin{\Theta(x, t)}\right]
\end{equation}
where $\Theta(x, t) = \theta(x-Vt) - \omega t$, and $V$ and $\omega \neq 0$ are constant parameters.
Such solutions were obtained for the integrable vector mKdV equation (\ref{Eq02}) in Ref. \cite{Karney1979};
one of the examples is shown in Fig. \ref{f02} for $V = -104$ and $\omega = 480$.
For the non-integrable mKdV equation (\ref{Eq01}) separation of variables and integration of
equations for $r$ and $\theta$ are less obvious if $\omega \neq 0$,
and the existence theory for helical solitons was not developed thus far.

The purpose of this paper is to apply the dynamical system methods to construct
the helical solitons in the vector non-integrable mKdV equation (\ref{Eq01}). We use the blow-up technique to unfold the singularity of the system of differential equations
for $(r,\theta)$ and construct smooth invariant manifolds from the critical point
representing the zero equilibrium. Continuation of the unstable manifold
numerically by means of an efficient algorithm enables us to detect existence of helical solitons in the parameter
plane $(V,\omega)$. The helical solitons appear as a result of co-dimension one bifurcation
and exist along a curve on the parameter plane $(V,\omega)$. From these numerical results, we conclude that
the helical solitons do exist for every $V < 0$ with a prescribed value of $\omega > 0$.
Moreover, a helical soliton with the opposite helicity can be constructed for $\omega < 0$ by a symmetry transformation.
In contrast to that, the helical solitons in the vector integrable mKdV equation (\ref{Eq02})
can exist in a certain two-dimensional region in the parameter plane with both positive
and negative values of $V$.

The paper is organized as follows. The main part of this paper is Section \ref{Sect2},
where the dynamical system methods are adopted for construction of helical solitons in Eq.~(\ref{Eq01}). Comparison with the helical solitons
in the vector integrable mKdV equation (\ref{Eq02}) is presented in Section \ref{Sect3}.
Direct numerical simulations indicating stability of helical solitons are described in Section
\ref{Sect5}. The concluding Section \ref{Sect4} contains the summary of this work.

Here we will investigate existence of helical solitons. It is convenient to combine
both models of  Eqs. (\ref{Eq01}) and (\ref{Eq02}) together into one equation with the parameter $\gamma$:
\begin{equation}
	\label{Eq05} %
	{\bf u}_t + \gamma|{\bf u}|^2 {\bf u}_x + (1-\gamma) (|{\bf u}|^2
	{\bf u})_x + {\bf u}_{xxx} = 0.
\end{equation}
When $\gamma = 0$, this equation reduces to the non-integrable vector mKdV equation (\ref{Eq01}),
whereas when $\gamma = 1$, this equation reduces to the integrable vector mKdV equation (\ref{Eq02}).

Using the polar form of vector ${\bf u} = (u_1, u_2) = r (\cos{\Theta}, \sin{\Theta})$ in Eq. (\ref{Eq05}),
we can obtain a set of two equations for $r(x, t)$ and $\Theta(x, t)$:
\begin{eqnarray}
	r_t + (3 - 2\gamma) r^2 r_x + r_{xxx} - 3 r_x (\Theta_x)^2 - 3r\Theta_x\Theta_{xx} & = & 0, \label{eq1} \\
	r\left(\Theta_t + r^2\Theta_x\right) + 3r_{xx}\Theta_x + 3 r_x\Theta_{xx} + r\Theta_{xxx} - r(\Theta_x)^3 & = & 0.
	\label{eq2}
\end{eqnarray}
The first equation (\ref{eq1}) can be written in the divergent
form for the variables $\rho = r^2$ and $w = \Theta_x$:
\begin{eqnarray}
	\label{eq3} %
	\rho_t + (3-2\gamma) \rho \rho_x + \left[\rho_{xx} - \frac{3}{4
		\rho} (\rho_x)^2 - 3 \rho w^2 \right]_x = 0.
\end{eqnarray}

For helical solitary solutions vanishing at the infinity and propagating with a constant speed $V$, we can assume that
$\rho(x,t) = \rho(x-Vt)$. Integrating then Eq. (\ref{eq3}) with the zero boundary conditions at the infinity, we obtain the following second-order ODE:
\begin{equation}
	\label{eq-rho} %
	\frac{d^2 \rho}{dx^2} - \frac{3}{4 \rho} \left( \frac{d\rho}{dx} \right)^2 - 3 \rho w^2 +
	\left( \frac{3}{2}  - \gamma \right) \rho^2 - V\rho = 0.
\end{equation}
This equation should be augmented by another equation derived from Eq. (\ref{eq2}).
Assuming that $\Theta(x,t) = \theta(x-Vt) - \omega t$, where $\omega$ is a constant frequency and noticing that $w = \Theta_x = \theta_x$, we obtain the following second-order ODE:
\begin{equation}
	\label{eq-w} %
	\frac{d^2 w}{dx^2} - w^3 + \frac{3}{2 \rho} w \frac{d^2 \rho}{d x^2} - \frac{3}{4\rho^2} w
	\left( \frac{d \rho}{dx} \right)^2 + \frac{3}{2\rho} \frac{d \rho}{dx} \frac{dw}{dx} + (\rho -
	V) w - \omega = 0.
\end{equation}

Helical solitons are defined to be solutions $(\rho,w)$ to the system (\ref{eq-rho}) and (\ref{eq-w})
with parameters $(V,\omega)$ satisfying the boundary conditions $(\rho,w) \to (0,k)$ as $|x| \to \infty$,
where $k$ is a real constant parameter. Assuming that solitons have exponential asymptotics at the infinity, i.e. function $\rho(x) \sim e^{\mp 2 \lambda x}$ as $x \to \pm \infty$, we obtain another constant parameter $\lambda > 0$. Substitution of  asymptotic expressions
for $\rho$ and $w$ into the system (\ref{eq-rho}) and (\ref{eq-w}) yields the following relationships between the parameters $(\lambda,k)$ and $(V,\omega)$:
\begin{eqnarray}
	V &=& \lambda^2 - 3 k^2, \label{lambda} \\
	\omega &=& 2 k (\lambda^2 + k^2). \label{omega}
\end{eqnarray}
The relationships (\ref{lambda}) and (\ref{omega}) define a mapping of the half-plane $\mathbb{R}^+ \times \mathbb{R}$
for $(\lambda,k)$ to a certain region in the parameter plane $(V,\omega)$. The region is located
to the right of the curves:
\begin{equation}
	\label{boundary}
	\omega =  \pm  2 \left( \frac{|V|}{3} \right)^{3/2}, \quad V < 0.
\end{equation}
For any point $(V,\omega)$ inside this region, we have $\lambda > 0$, whereas the boundary curves correspond to $\lambda = 0$.

If a solution of system (\ref{eq-rho}) and (\ref{eq-w}) is found for some $(\lambda,k)$ and corresponding values of $(V,\omega)$, then another symmetric solution can be obtained for $(\lambda,-k)$ and $(V,-\omega)$ by the reflection symmetry $\rho \mapsto \rho$ and $w \mapsto -w$ in the system (\ref{eq-rho}) and (\ref{eq-w}). Such pair of solutions would correspond to solitons of opposite helicity, when vector $\bf u$ rotates either clockwise or couterclockwise in the process of propagation along the $x$-axis.

When $k \to 0$ and $\omega \to 0$, the helical solitons (\ref{helic-sol})
reduce to the traveling solitons (\ref{trav-sol}) with (\ref{trav-sol-polar})
and $V = \lambda^2 > 0$. As has been mentioned in the introduction, such traveling solitons exist for every $V > 0$ (if $\gamma < 3/2$),
because $w = 0$ is the invariant reduction of Eq.~(\ref{eq-w}) with $\omega = 0$,
after which soliton solutions to Eq.~(\ref{eq-rho}) can be constructed in the exact form for $V > 0$.
It is, however, beyond the scopes of this work to consider other possible solutions to system (\ref{eq-rho}) and (\ref{eq-w}) with $\omega = 0$.

\subsection{Reformulation of the problem and the blow-up scaling}
\label{Subsec2.1}

To construct soliton solutions it os convenient to reformulate the set of two second-order equations, (\ref{eq-rho}) and
(\ref{eq-w}), in terms of a dynamical system in the spatial coordinate $x$. Note that according to the product rule, we have:
$$
\frac{d^2}{dx^2} (\rho^{3/4} w) = \rho^{3/4} \frac{d^2
	w}{dx^2} + \frac{3}{2} \rho^{-1/4} \frac{d \rho}{dx} \frac{dw}{dx}
+ \frac{3}{4} \rho^{-1/4} w \frac{d^2 \rho}{dx^2} - \frac{3}{16}
\rho^{-5/4} w \left( \frac{d \rho}{dx} \right)^2.
$$

Application of this formula in Eq.~(\ref{eq-w}) multiplied by $\rho^{3/4}$ gives:
$$
\frac{d^2}{dx^2} (\rho^{3/4} w) + \frac{3}{4} \rho^{-1/4} w \left[
\frac{d^2 \rho}{d x^2} - \frac{3}{4 \rho} \left(\frac{d \rho}{dx}
\right)^2 \right] + \left(\rho w - Vw - w^3  - \omega\right)\rho^{3/4} = 0.
$$
Elimination the derivatives of $\rho_x$ with the help of Eq.~(\ref{eq-rho}) reduces
this equation to the final form:
\begin{equation} %
	\label{Eq08}%
	\frac{d^2}{dx^2} (\rho^{3/4} w) = \frac 14\left[\left(\frac{1}{2}
	- 3\gamma\right)\rho w + Vw - 5w^3 + 4\omega \right]\rho^{3/4}.
\end{equation}
On the other hand, Eq.~(\ref{eq-rho}) multiplied by $\rho^{-3/4}/4$ takes the form:
\begin{equation} %
	\label{Eq09}%
	\frac{d^2}{dx^2} (\rho^{1/4}) = \frac 14\left[V + 3w^2 -
	\left(\frac 32 -\gamma\right) \rho\right]\rho^{1/4}.
\end{equation}

Introducing new variables $\varphi = \rho^{1/4}$, $\psi =
\rho^{3/4} w$, and $\tau = x/2$, we can rewrite
Eqs. (\ref{Eq08}) and (\ref{Eq09}) in the equivalent form:
\begin{eqnarray}
	\frac{d^2 \varphi}{d \tau^2} &=& 3\psi^2 \varphi^{-5} -
	\left(\frac{3}{2} - \gamma\right)\varphi^5 + V\varphi, \label{eq-u} \\
	\frac{d^2\psi}{d \tau^2} &=&  - 5\psi^3\varphi^{-6} + \left(\frac{1}{2}-3\gamma\right)
	\varphi^4 \psi + V\psi + 4 \omega \varphi^3.
	\label{eq-v}
\end{eqnarray}

The asymptotic behaviour of new variables at the infinity follows from
the corresponding asymptotics of functions $\rho$ and $w$:
$$
\varphi(\tau) \sim e^{\pm \lambda \tau}, \quad
\psi(\tau) \sim -k e^{\pm 3 \lambda \tau}, \quad \mbox{\rm as} \quad
\tau \to \mp \infty.
$$
Substitution of these representations into the system (\ref{eq-u})--(\ref{eq-v})
provides the correct formulae (\ref{lambda}) and (\ref{omega}) for $V$ and $\omega$.

The spatial dynamical system (\ref{eq-u}) and (\ref{eq-v}) can be rewritten
as the set of four first-order ODEs:
\begin{equation}
\label{dyn-sys} %
\frac{d}{d\tau} \left[\begin{array}{c} \varphi \\ \eta \\ \psi
\\ \zeta \end{array} \right] =
\left[ \begin{array}{c} \eta \\ 3\psi^2 \varphi^{-5} - \left(\frac{3}{2}-\gamma \right) \varphi^5 + V\varphi \\
\zeta \\ - 5 \psi^3 \varphi^{-6} + \left( \frac{1}{2} - 3\gamma \right) \varphi^4 \psi + V\psi + 4 \omega\varphi^3 \end{array} \right].
\end{equation}

This dynamical system is singular at $\varphi = 0$. Inspired by
the linearized behavior of variable $(\varphi, \psi)$ in the
vicinity of the point $(0, 0)$, we can unfold the singularity by
means of the following elementary transformation:
$$
\left[ \begin{array}{c} \varphi \\ \eta \\ \psi \\ \zeta
\end{array} \right] = \left[\begin{array}{c} \Phi \\ \Phi Y \\ \Phi^3 \Psi \\
\Phi^3 Z \end{array} \right],
$$
where the new variables $(\Phi,Y,\Psi,Z)$ obey the equivalent dynamical system:
\begin{equation}
\label{dyn-sys-scaled} %
\frac{d}{d\tau} \left[\begin{array}{c} \Phi \\
Y \\ \Psi \\ Z \end{array} \right] =
\left[\begin{array}{c} \Phi Y \\ - Y^2 + 3\Psi^2 - \left( \frac{3}{2} - \gamma \right) \Phi^4 + V \\
Z - 3\Psi Y \\  - 3ZY - 5\Psi^3 + \left( \frac{1}{2} - 3\gamma \right) \Phi^4 \Psi + V\Psi + 4\omega\end{array} \right].
\end{equation}
The rescaled set of equations (\ref{dyn-sys-scaled}) is no longer singular at $\Phi = 0$.

\subsection{Critical points and invariant manifolds}
\label{Subsec2.2}

The dynamical system (\ref{dyn-sys-scaled}) has
two-parametric family of critical points:
\begin{equation}
	\label{crit-points}
	(\Phi,Y,\Psi,Z) = (0,Y_0,\Psi_0,Z_0),
\end{equation}
where $Z_0 = 3\Psi_0 Y_0$, and $Y_0$ and $\Psi_0$ are related with the parameters $V$ and
$\omega$ by the following algebraic equations:
\begin{eqnarray}
	V &=& Y_0^2 - 3\Psi_0^2, \label{parameterization-c-omega1} \\
	\omega &=& 2 \Psi_0 (Y_0^2 + \Psi_0^2).
	\label{parameterization-c-omega2}
\end{eqnarray}
This algebraic system coincides with the system (\ref{lambda})--(\ref{omega}) if we set $Y_0 = \lambda$ and
$\Psi_0 = k$. The transformation $(Y_0,\Psi_0) \mapsto (V, \omega)$ is
invertible for every real $Y_0 > 0$ and $\Psi_0$, because its Jacobian is nonzero:
$$
\frac{\partial (V, \omega)}{\partial (Y_0, \Psi_0)} = 4Y_0(Y_0^2 + 9\Psi_0^2) \neq 0.
$$

The matrix of the linearised system in the vicinity of the critical point is:
$$
\left[\begin{array}{cccc} %
Y_0 & 0 & 0 & 0 \\
0 & -2Y_0 & 6 \Psi_0 & 0 \\
0 & -3\Psi_0 & -3 Y_0 & 1 \\
0 & -3 Z_0 & V -15\Psi_0^2 & - 3Y_0 \end{array} \right]
$$
One of the eigenvalues of this matrix is $\mu_1 = Y_0$. The other
three eigenvalues can be found from the following cubic equation: %
\begin{equation}%
	\label{Eq10} %
	\mu^3 + 8 Y_0 \mu^2 + 4 (5 Y_0^2 + 9 \Psi_0^2) \mu + 16 Y_0 (Y_0^2
	+ 9\Psi_0^2) = 0,
\end{equation}
where we have used that $Z_0 = 3\Psi_0 Y_0$ and $V = Y_0^2 - 3
\Psi_0^2$. The roots of this equation can be easily found if we notice that one of them is real $\mu_2 = -4 Y_0$; then the other two roots are complex-conjugate:
\begin{equation}%
	\label{Eq11} %
	\mu_{3,4} = -2Y_0 \pm 6 i \Psi_0.
\end{equation}

The remarkable property of linearization for $Y_0 > 0$ is that
only one root $\mu_1$ is positive, whereas the other three roots
have negative real parts. Therefore, there
exists a one-dimensional unstable curve in the phase space that originates
at the critical point $(0,Y_0,\Psi_0,Z_0)$ with $Z_0 = 3 Y_0
\Psi_0$ belonging to the two-parametric family of critical points (\ref{crit-points}).
Using parametrization (\ref{parameterization-c-omega1}) and (\ref{parameterization-c-omega2})
of $(V, \omega)$ in terms of $Y_0 = \lambda$ and
$\Psi_0 = k$, we can rewrite the dynamical system
(\ref{dyn-sys-scaled}) in the following equivalent form:
\begin{equation}
\label{dyn-sys-numerics} %
\frac{d}{d\tau} \left[ \begin{array}{c} \Phi \\ Y \\ \Psi \\ Z \end{array} \right] =
\left[\begin{array}{c} \Phi Y \\ \lambda^2 - Y^2 + 3( \Psi^2 - k^2) - (\frac{3}{2}-\gamma)\Phi^4 \\
- 3\Psi Y + Z \\ (\lambda^2 - 3 k^2  - 5\Psi^2)\Psi + (\frac{1}{2}-3\gamma)\Phi^4 \Psi  + 8 k (\lambda^2 + k^2) - 3ZY
\end{array} \right].
\end{equation}
This representation allows us to construct a numerical shooting algorithm based on the approximation of the
trajectory along the unstable curve. When the unstable curve intersects with the plane
$Y = Z = 0$ at a certain instant $\tau_0$ of the ``time'' $\tau$, the calculation can be terminated,
and the trajectory in phase space can be continued beyond $\tau_0$, as the even functions for $\Phi$, $\Psi$
and odd functions for $Y$, $Z$ in variable $\tau$. In terms of original variables $\rho$, $w$
as functions of $x$ such solution corresponds
to the helical soliton vanishing at the infinities and symmetric with respect to its center.

\subsection{Numerical solutions on the basis of the shooting algorithm}
\label{Subsect2.3}

\begin{figure}[h]
	\centerline{\includegraphics [scale=0.55]{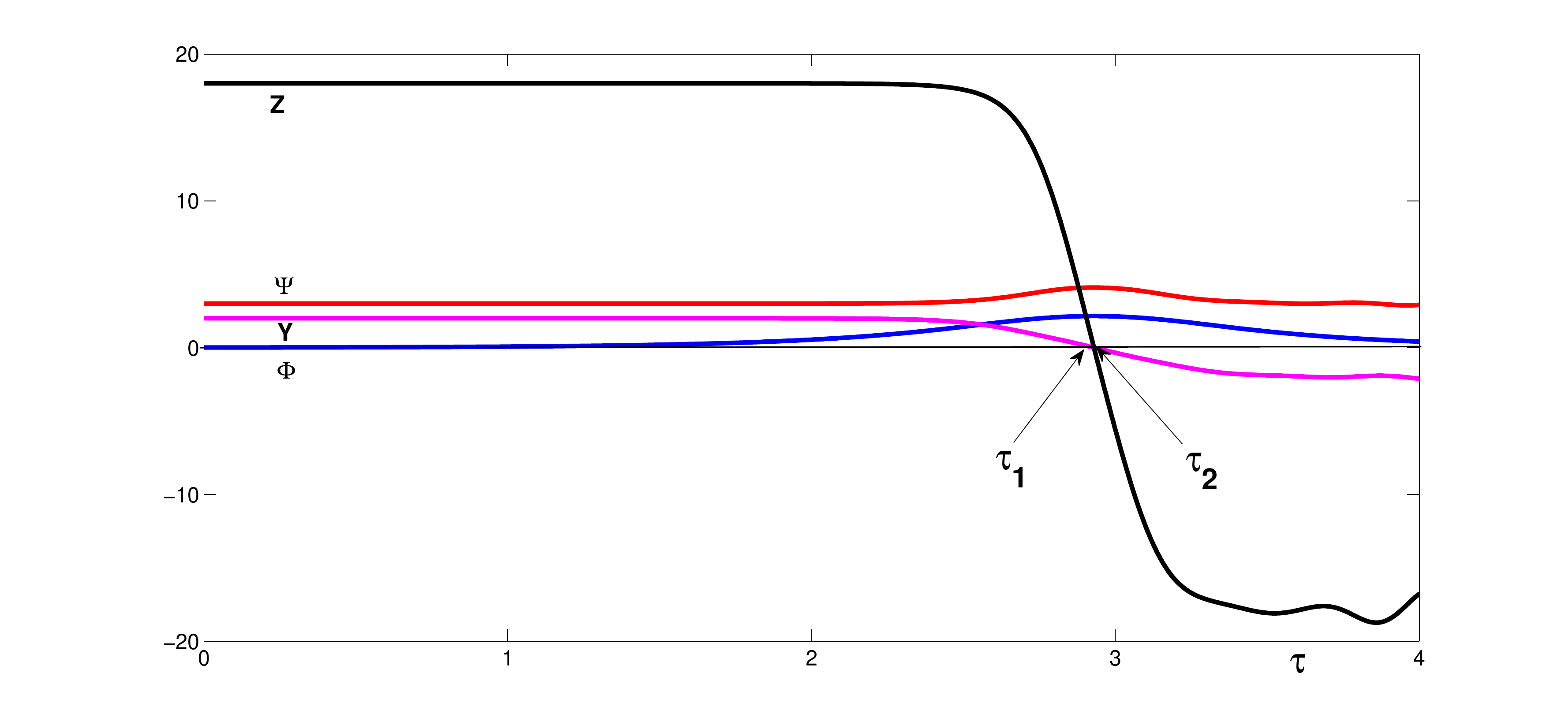}}
	\caption{Numerical solution to the dynamical system (\ref{dyn-sys-numerics})
		corresponding to the unstable curve with $\varepsilon = 0.01$, $\lambda = 2$, and $k = 3$.}
	\label{fig-traj}
\end{figure}

\begin{figure}[h]
	\centerline{\includegraphics [scale=0.55]{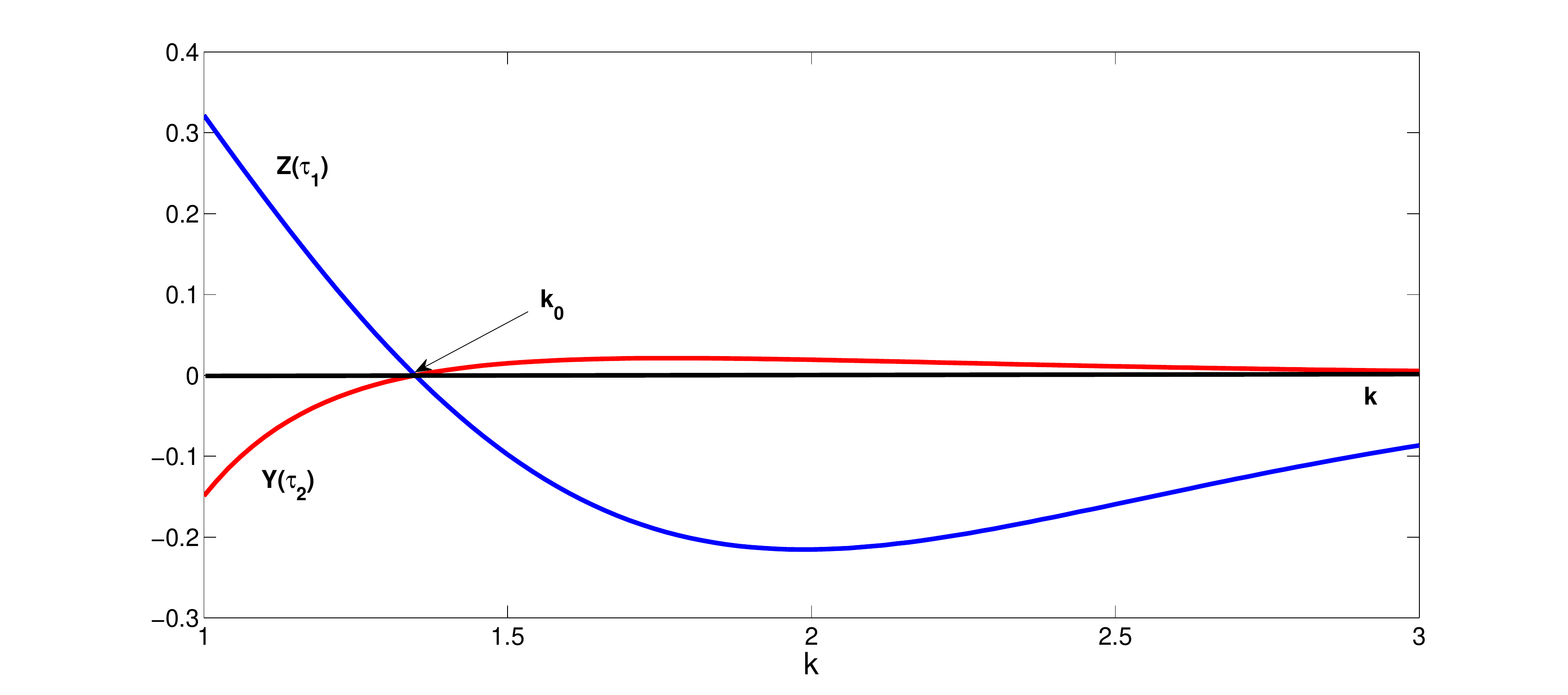}}
	\caption{The dependencies of $Z(\tau_1)$ at $Y (\tau_1)= 0$ (blue) and $Y(\tau_2)$ at $Z(\tau_2) = 0$ (red)
		on the parameter $k$ for $\lambda = 2$.}%
	\label{fig-intersection}
\end{figure}

Let us consider the non-integrable vector mKdV equation (\ref{Eq01}) which corresponds to $\gamma = 0$ in Eq.~(\ref{dyn-sys-numerics}). If we solve the set of ODEs (\ref{dyn-sys-numerics}) for $(\Phi,Y,\Psi,Z)$ subject to the initial condition
$(\varepsilon,\lambda,k,3\lambda k)$, where $\varepsilon \ll 1$, $\lambda > 0$, and any $k$,
we can find the instances of $\tau$ when $Y$ and $Z$ turn to zero.
Fig. \ref{fig-traj} illustrates that the intersections of these functions with
the $\tau$-axis occur generally at different instances, say $\tau_1$ and $\tau_2$.
However, varying the parameter $k$ with fixed $\lambda$, we can find
a value of $k$, say $k_0$, when intersections occur at the same instants of $\tau$, i.e.,
when $\tau_1 = \tau_2$, see Fig. \ref{fig-intersection}.

Blue line in Fig. \ref{fig-intersection} shows the dependence of $Z(\tau_1)$ on the parameter $k$ at
the first instance of time $\tau_1$ when $Y(\tau_1) = 0$, and red line shows the dependence of
$Y(\tau_2)$ on the parameter $k$ at the first instance of time $\tau_2$ when $Z(\tau_2) = 0$. At the point $k_0$ where these two curves intersect, we have $\tau_1 = \tau_2$, and the unstable curve originated at the critical point hits the plane $Y = Z = 0$. For such values of $k$
and $\lambda$ the trajectory can be symmetrically continued to form a solution for the helical soliton as explained above.

\begin{figure}[b!]
	\centerline{\includegraphics [scale=0.55]{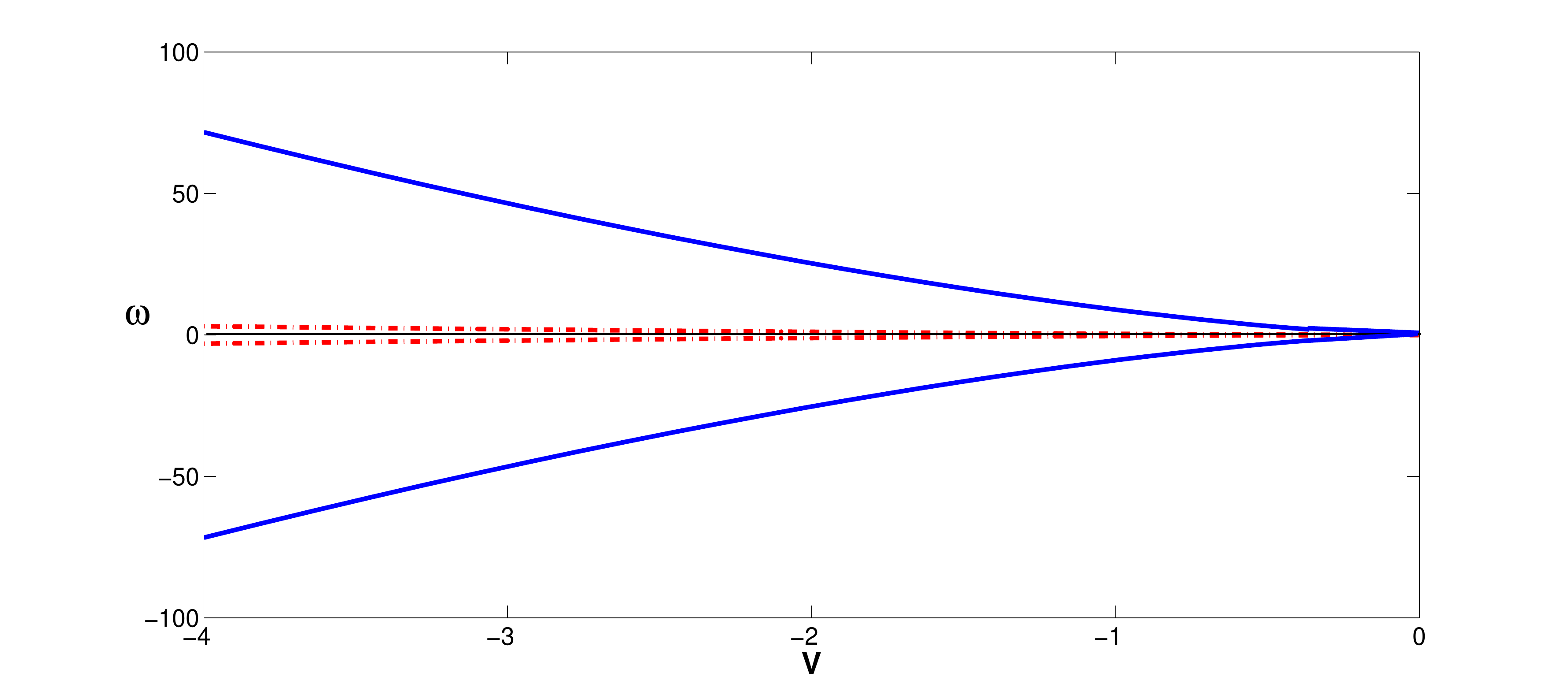}}
	\caption{The existence curve for helical solitons in the
		non-integrable vector mKdV equation (\ref{Eq01}) on the
		parameter plane $(V, \omega)$ is shown by solid blue line whereas the
		boundary of the admissible region (\ref{boundary}) is shown by dash-dotted red line.}
	\label{fig-parameter-plane}
\end{figure}

Varying $\lambda$ and detecting $k_0$ for each $\lambda$ gives the existence curve
in the parameter plane $(\lambda,k)$. Using the parameterisations (\ref{lambda}) and
(\ref{omega}), we can determine the corresponding values of $(V,\omega)$ from $(\lambda,k)$.
This allows us to plot the existence curve for the helical solitons in the parameter plane $(V, \omega)$;
such curves are shown by solid blue line in Fig. \ref{fig-parameter-plane}. Dash-dotted red line shows the boundary (\ref{boundary})
of the admissible domain for existence of helical solitons.

As we can see, the existence curves originate at the point $(V,\omega) = (0,0)$ and are located in the region where $V < 0$ in agreement with Eq.~(\ref{boundary}). Therefore,
we conclude that the helical solitons (\ref{helic-sol}) in the
non-integrable vector mKdV equation (\ref{Eq01}) do exist as a result of the co-dimension one
bifurcation, and their velocities are negative in contrast to the planar travelling solitons (\ref{trav-sol})
with (\ref{trav-sol-polar}) and $V > 0$.

\begin{figure}[t!]
{\centerline{\includegraphics [scale=0.35]{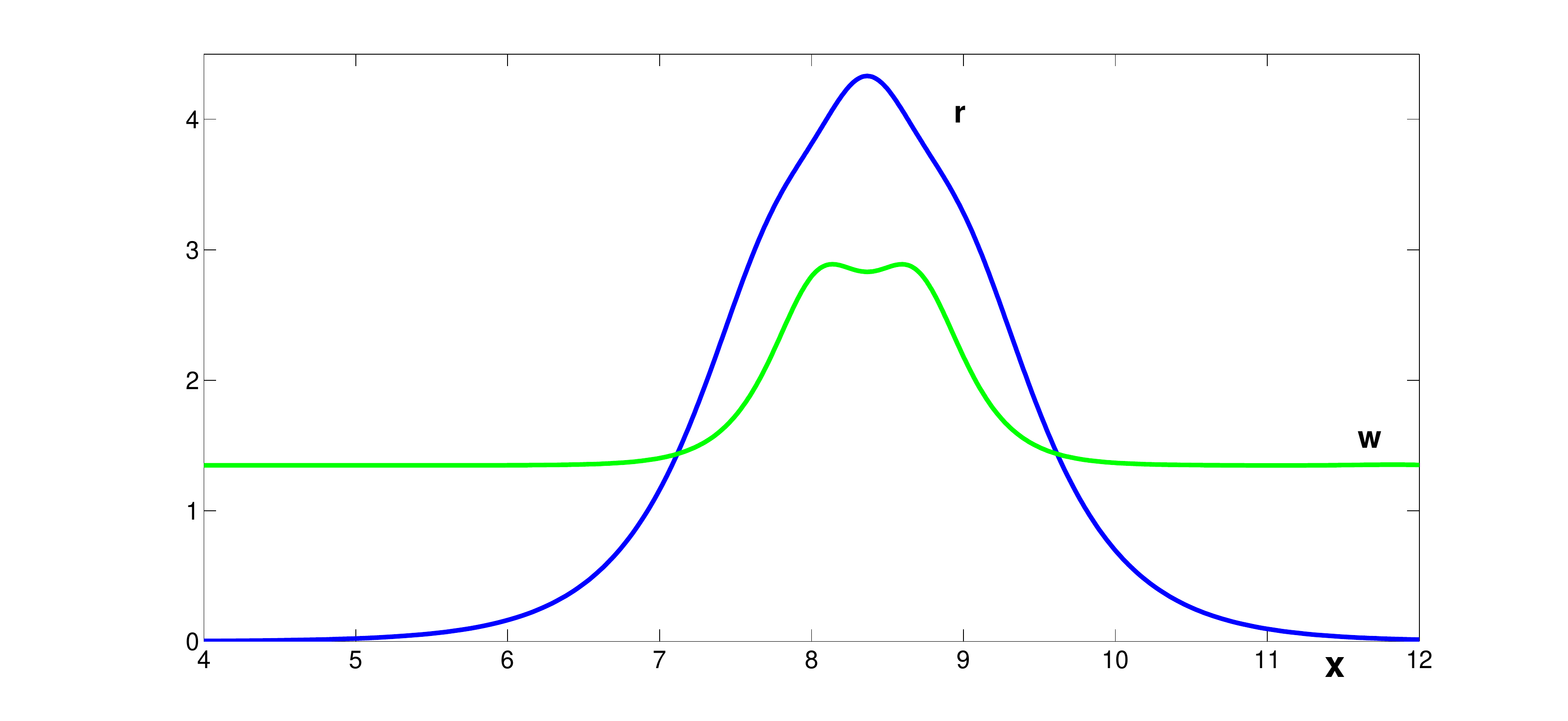}}}
{\centerline{\includegraphics [scale=0.35]{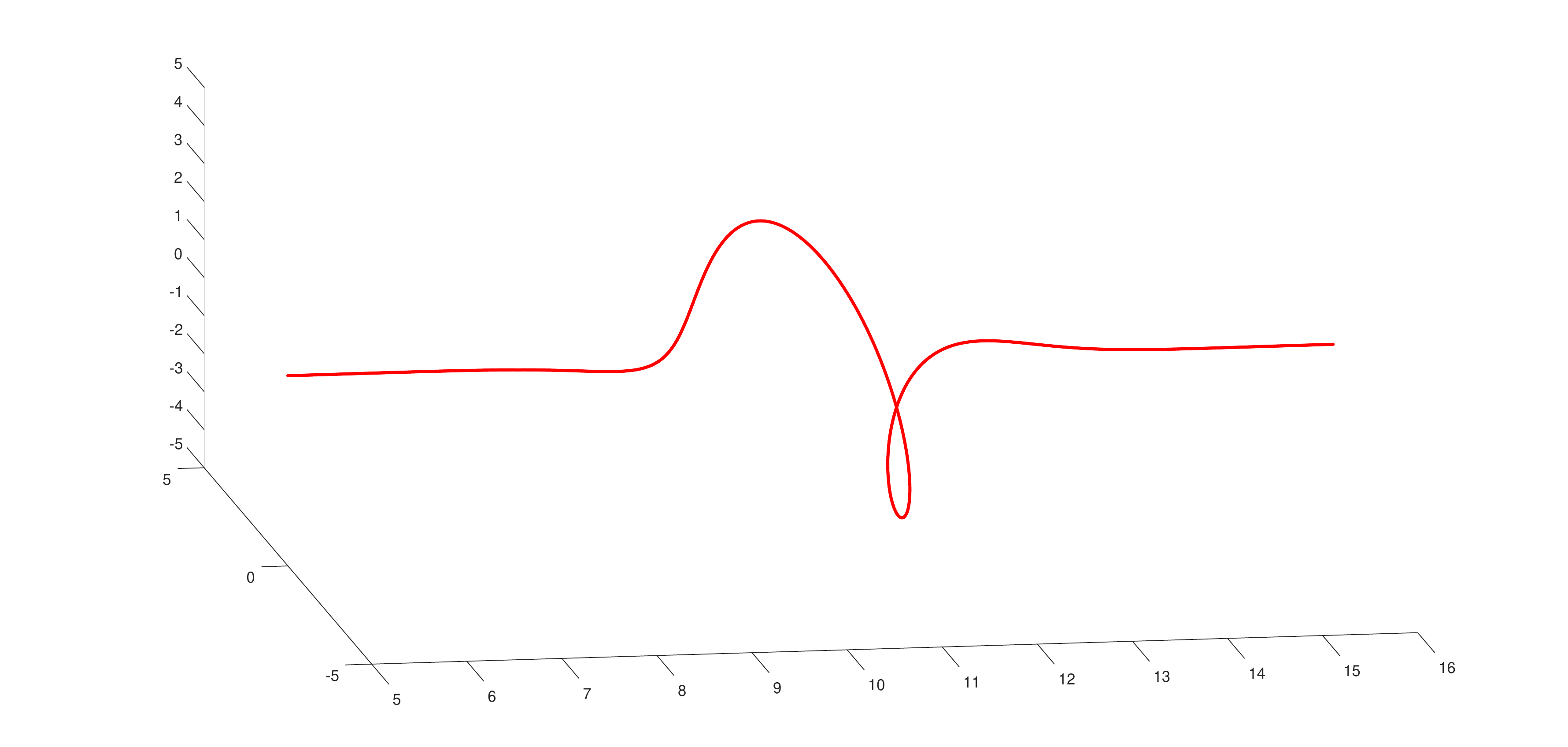}}}
	\begin{picture}(300,6)%
	\put(5,250){\large a)}%
	\put(5,90){\large b)}%
	\put(205,23){$x$}%
	\put(38,55){$y$}%
	\put(23,135){$z$}%
	\end{picture}
	\caption{A typical profile of the helical soliton in the non-integrable vector mKdV equation
		(\ref{Eq01}) with $V \approx -1.45$ and $\omega \approx 15.69$. Panel (a) shows the solution in variables
		$(r,w)$ as functions of $x$. Panel (b) displays the solution in variables $(u_1,u_2)$ in the three-dimensional space.} %
	\label{fig-soliton}
\end{figure}

The first intersections of the unstable curves with the plane $Y=Z=0$ provide
the strictly positive profile for $\Phi$ in $\tau$. This yields the helical solitons
with a single-humped profile for $r$ in $x$.
Figure \ref{fig-soliton} shows a typical profile of the helical soliton for $\lambda = 2$ and $k \approx 1.348$ which corresponds to $V \approx -1.45$ and $\omega \approx 15.69$.

\section{Helical solutions in the integrable vector mKdV equation}%
\label{Sect3}

The integrable vector mKdV equation (\ref{Eq02}) corresponds to Eq. (\ref{Eq05}) with $\gamma = 1$. 
This equation can be used as the limiting case for the non-integrable equation (\ref{Eq01}) and 
due to its integrability, can provide a certain insight about possible solutions.

As has been noticed in Ref. \cite{Karney1979}, Eq. (\ref{Eq01}) can be rewritten in one of the following equivalent forms:
\begin{eqnarray}
	u_t + |u|^2u_x + u_{xxx} &=& -u\left(|u|^2\right)_x,
	\label{EqA1} \\%
	u_t + 3|u|^2u_x + u_{xxx} &=& 2iu|u|^2\Theta_x.
	\label{EqA2}
\end{eqnarray}
where $u = u_1 + i u_2 = r\,e^{i\Theta}$ is a function of $x$ and $t$.
It follows from these representations that the right-hand
sides of equations are negligibly small in the following two cases:

\begin{enumerate}
	\item If a spatial scale of variation of $r = |u|$
	is much greater than the spatial scale of variation
	of $\Theta = {\rm arg}(u)$, i.e. $|r_x | \ll |r \Theta_x|$.
	In this case, Eq.~(\ref{EqA1}) reduces to the integrable Eq.~(\ref{Eq02}).
	
	\item If a spatial scale of variation of $\Theta = {\rm arg}(u)$
	is much greater than the spatial scale of variation of $r = |u|$,
	i.e. $|r \Theta_x| \ll |r_x|$. In this case,
	Eq.~(\ref{EqA2}) also reduces to the integrable Eq.~(\ref{Eq02}) with
	the following change: the nonlinear coefficient is three times bigger.
\end{enumerate}


Let us now inspect how the dynamical system methods apply to the integrable vector mKdV equation (\ref{Eq02}). 
If $\gamma = 1$, the system (\ref{eq-rho}) and (\ref{eq-w}) with $w = k = const$ can be reduced to the scalar equation for $\rho$. 
Indeed, eliminating $\rho_{xx}$ from  the system (\ref{eq-rho}) and (\ref{eq-w})
with $w = k$, we obtain:
\begin{equation}%
	\label{Eq12}%
	\frac{3}{2}\left(\frac{d \rho}{d x} \right)^2 + \rho^2 \left(\rho
	+ 14k^2 + 2V - 4\frac{\omega}{k} \right) = 0.
\end{equation}
Substitution of this into Eq.~(\ref{eq-rho}) yields:
\begin{equation}%
	\label{Eq13}%
	\frac{d^2 \rho}{d x^2} + \rho^2 + \left(4k^2 - 2\frac{\omega}{k}\right) \rho = 0.
\end{equation}
After integration of this equation with the zero boundary conditions at the
infinity, we obtain:
\begin{equation}%
	\label{Eq14}%
	\frac{3}{2}\left(\frac{d \rho}{d x} \right)^2 + \rho^2 \left(\rho
	+ 6k^2 - 3\frac{\omega}{k} \right) = 0.
\end{equation}
Equations (\ref{Eq12}) and (\ref{Eq14}) are compatible if and only
if $\omega = 2k(V + 4k^2)$, which agrees with Eqs. (\ref{lambda})
and (\ref{omega}). From Eq. (\ref{Eq13}) it also follows that if a
solution has the exponential asymptotics: $\rho(x) \sim e^{\mp 2 \lambda x}$
as $x \to \pm \infty$, then $\omega = 2k( \lambda^2 + k^2)$
in agreement with Eq.~(\ref{omega}).

When a trajectory of the system (\ref{dyn-sys-numerics}) with
$\gamma = 1$ is considered along the unstable curve
from the initial point $(\varepsilon,Y_0,\Psi_0,Z_0)$ with
$Z_0 = 3 Y_0\Psi_0$, $Y_0 = \lambda$, $\Psi_0 = k$,
and small $\varepsilon$, then $\Psi = k$ and $Z = 3kY$ are preserved in 
$\tau$ and the system (\ref{dyn-sys-numerics}) with $\gamma = 1$
reduces to the second-order differential equation:
\begin{equation}%
	\label{dyn-integrable} %
	\frac{d^2 \Phi}{d \tau^2} =  \lambda^2\Phi - \frac{1}{2}\Phi^5.
\end{equation}
Solution to this equation can be readily obtained in the explicit
form:
\begin{equation}%
	\label{Eq15} %
	\Phi(\tau) = (6\lambda^2)^{1/4} {\rm sech}^{1/2}(2 \lambda \tau),
\end{equation}
which yields the exact solution
\begin{equation}
	r(x) = \lambda \sqrt{6} {\rm sech}(\lambda x), \quad w(x) = k.
\end{equation}
Therefore, the helical soliton in the integrable case with $\gamma
= 1$ exists for every $(V, \omega)$ given by the image of
transformation $(\lambda, k) \mapsto (V, \omega)$ in
(\ref{lambda}) and (\ref{omega}) with $\lambda > 0$ and any real
$k$. In other words, it exists in the entire region in the parameter
plane $(V,\omega)$ to the right of the boundary (\ref{boundary})
including the region with positive speed $V$.

In terms of original variables the helical soliton is:
\begin{equation}%
	\label{Eq16} %
	{\bf u} = (u_1, u_2) = \lambda\sqrt{6}\,{\rm
		sech}\lambda\xi\, \left[ \cos(k\xi - \omega t + \theta_0), \;
	\sin(k\xi - \omega t + \theta_0) \right],
\end{equation}
where $\xi = x - Vt$ and $\theta_0$ is an arbitrary constant.
Such solution was derived by means of the inverse scattering method in
Ref. \cite{Karney1979} and presented in the complex form $u = u_1 + i u_2 =
\lambda\sqrt{6}\,{\rm sech}\lambda\xi e^{i(\omega t - k\xi +	\theta_0)}$.

\begin{figure}[h!]
{\centerline{\includegraphics [scale=0.35]{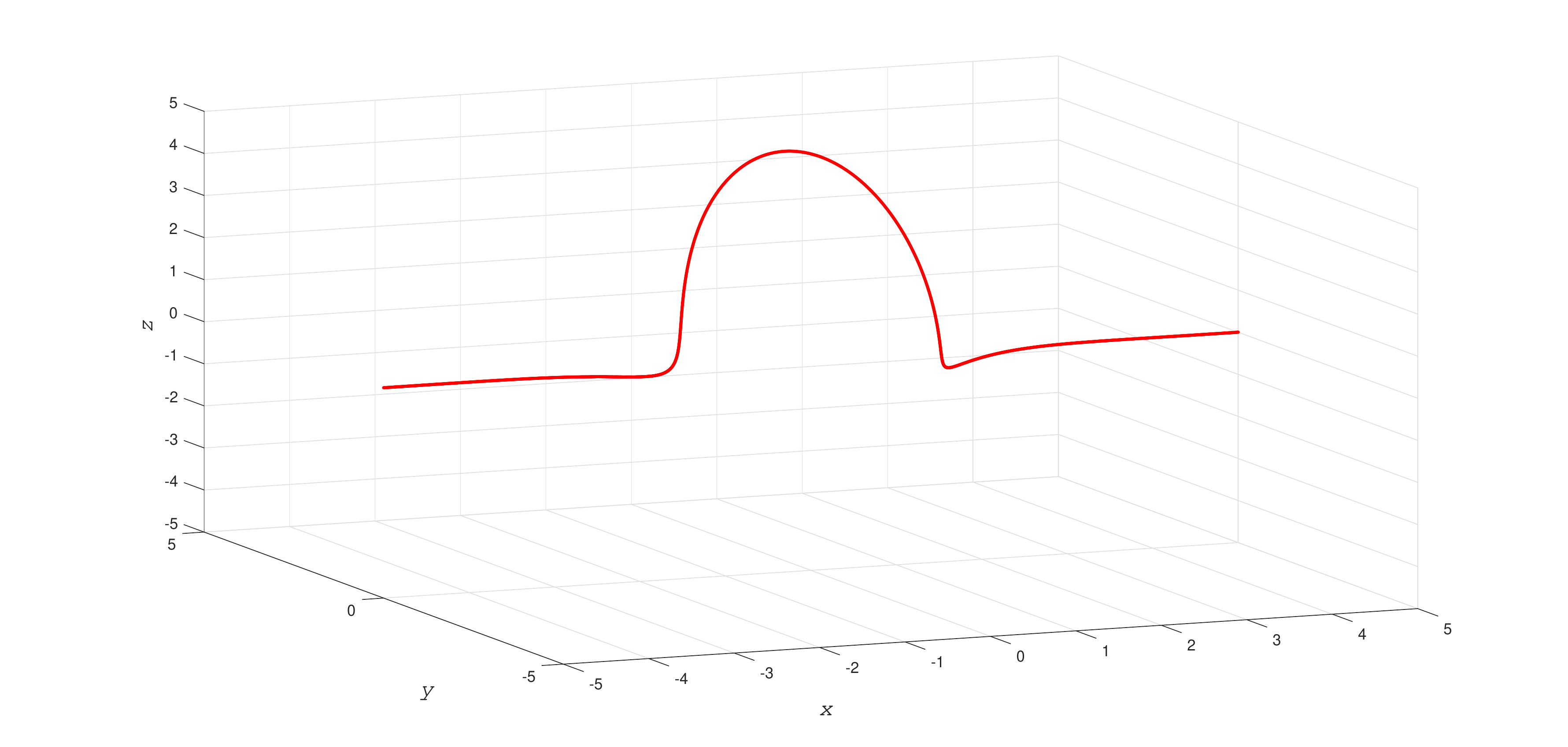}}}
{\centerline{\includegraphics [scale=0.35]{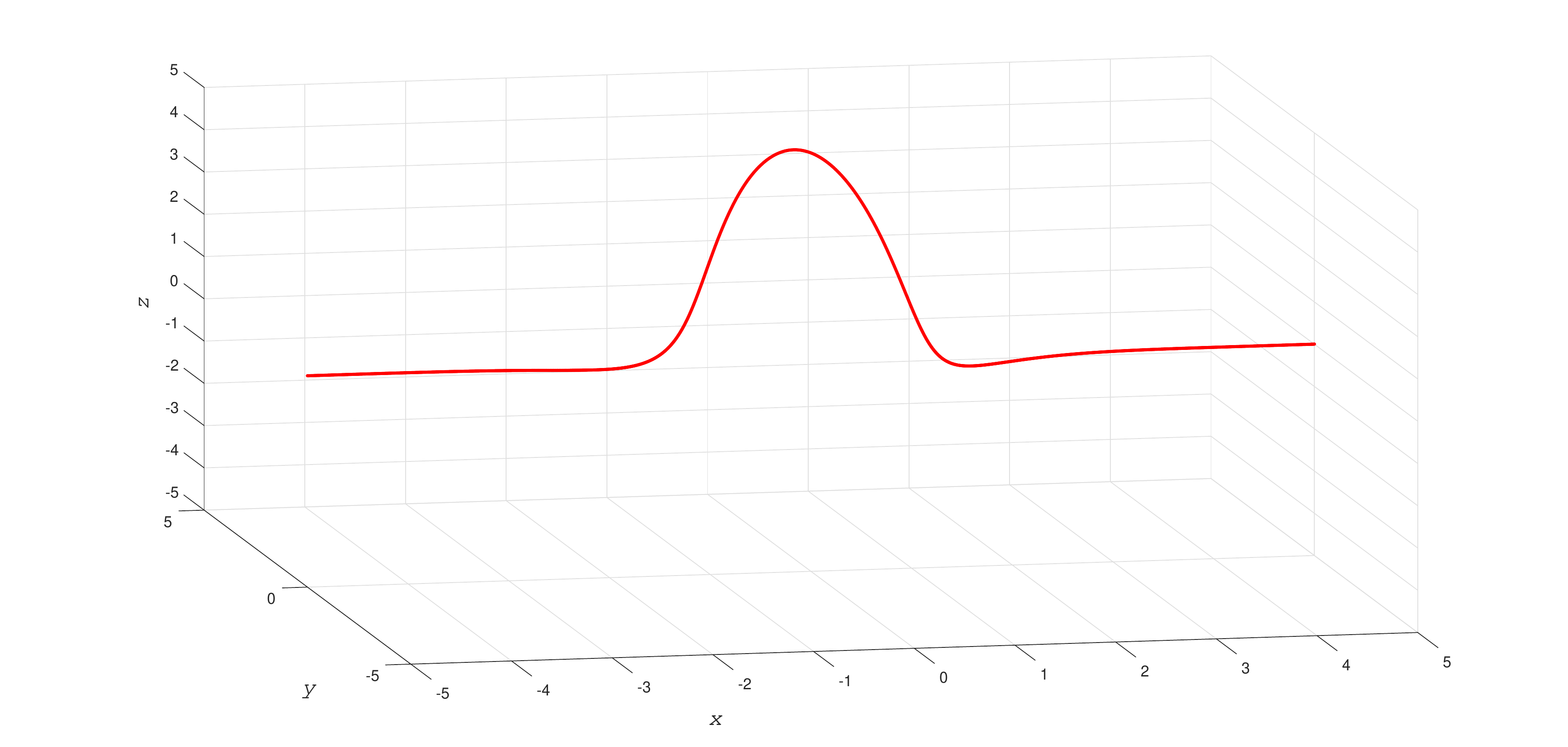}}}
	\begin{picture}(300,6)%
	\put(0,280){\Large $a)$}%
	\put(0,110){\Large $b)$}%
	\end{picture}
	\vspace{-1.0cm}%
	\caption{Two examples of helical solitons in the integrable vector
		mKdV equation (\ref{Eq02}) with $\lambda = 2$ and (a)
		$k_1 \approx 1.348$ as in Fig. \ref{fig-soliton} ($V \approx -1.451$, $\omega \approx 15.683$) and (b) $k_2 = \lambda/\sqrt{3} \approx 1.155$ ($V = 0$, $\omega \approx 12.317$).}%
	\label{f07}
\end{figure}

Examples of helical solitons are shown in Fig. \ref{f07} for
$\lambda = 2$ and two values of $k$: $k_1 \approx 1.348$ (frame a) and $k_2
= \lambda/\sqrt{3} \approx 1.155$ (frame b). The former case corresponds
to the same choice of parameters $\lambda$ and $k$ as in Fig. \ref{fig-soliton}, whereas the latter case
corresponds to the standing soliton with $V = 0$ according to Eq.~(\ref{lambda}).
One more examples of helical soliton was presented in Fig. \ref{f02}
for $\lambda = 2$, $k = 6$ which gives $V = -104$, $\omega = 480$.

In the particular case of $k = 0$ (and hence $\omega = 0$ as per Eq.~(\ref{omega})) the helical soliton (\ref{Eq16})
degenerates into a planar soliton (\ref{trav-sol}) with (\ref{trav-sol-polar}), where the parameter $\Theta = \theta_0$
represents the polarization angle in the $(y,z)$ plane.

\section{Stability of helical solitons}
\label{Sect5}

Stability of helical solitons was confirmed by direct numerical calculations of the vector mKdV equations (\ref{Eq01})
and (\ref{Eq02}). The accuracy of calculations was controlled through the conservation of the total energy
$I_2 = \int_{\mathbb{R}} |{\bf u}|^2dx$. This integral quantity was preserved in the numerical computations
with the relative error less than $10^{-2}$\%.

In the integrable case of Eq.~(\ref{Eq02}) it was confirmed that the initially given helical soliton simply propagates with a constant speed in accordance with the theoretical prediction. This, in particular, occurs with the helical soliton shown in Fig. \ref{f02}. When the same initial condition is used for the nonintegrable case of Eq.~(\ref{Eq01}), then the helical solitary wave becomes stationary after a short transient period, but gets a smaller amplitude.
The solitary wave is accompanied by a small non-stationary dispersive wavetrain of a helical structure.

\begin{figure}[h!]
	{\centerline{\includegraphics [scale=0.5]{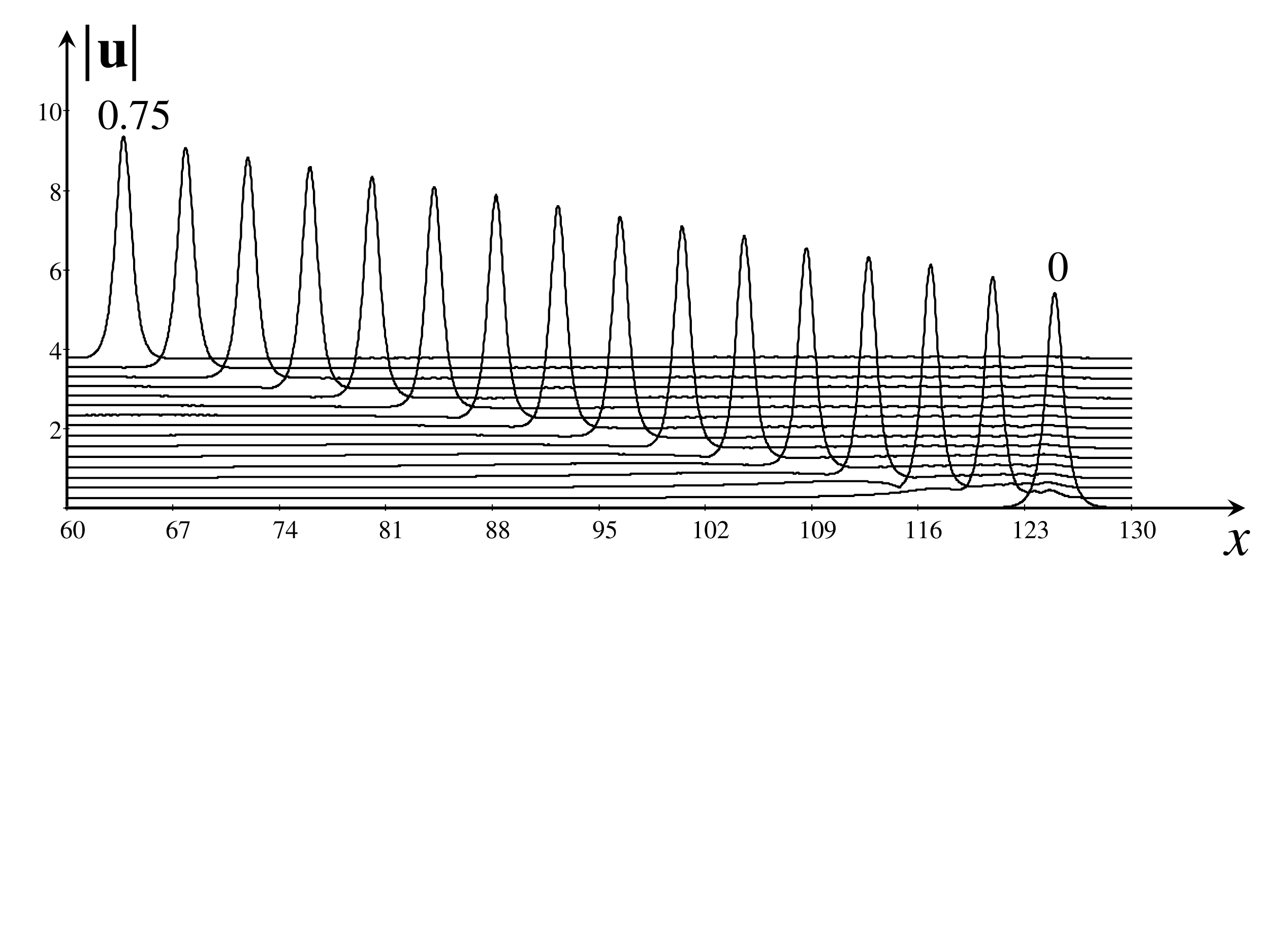}}}
	\vspace{-4.0cm}%
	\caption{Evolution of the helical pulse within the non-integrable Eq. (\ref{Eq01}) moving to the left and emitting a small-amplitude helical wavetrain. Numbers show the initial and ending time instants. Curves were plotted with the time step 0.05.}%
	\label{f08}
\end{figure}

A similar phenomenon was observed within the non-integrable Eq. (\ref{Eq01}) when the initial amplitude of a helical pulse was 10\% greater than the amplitude of a helical soliton shown in Fig. \ref{f02}. The helical soliton moved with the negative velocity behind the emitted small-amplitude helical wavetrain. Its amplitude slightly increased and stabilized then at a certain level; this evolution is illustrated by Fig. \ref{f08} where we show the modulus of vector $\bf u$ at several instants of time, from $t = 0$ to $t = 0.75$ with the time step $\Delta t = 0.05$. 

\begin{figure}[h!]
	{\centerline{\includegraphics [scale=0.35]{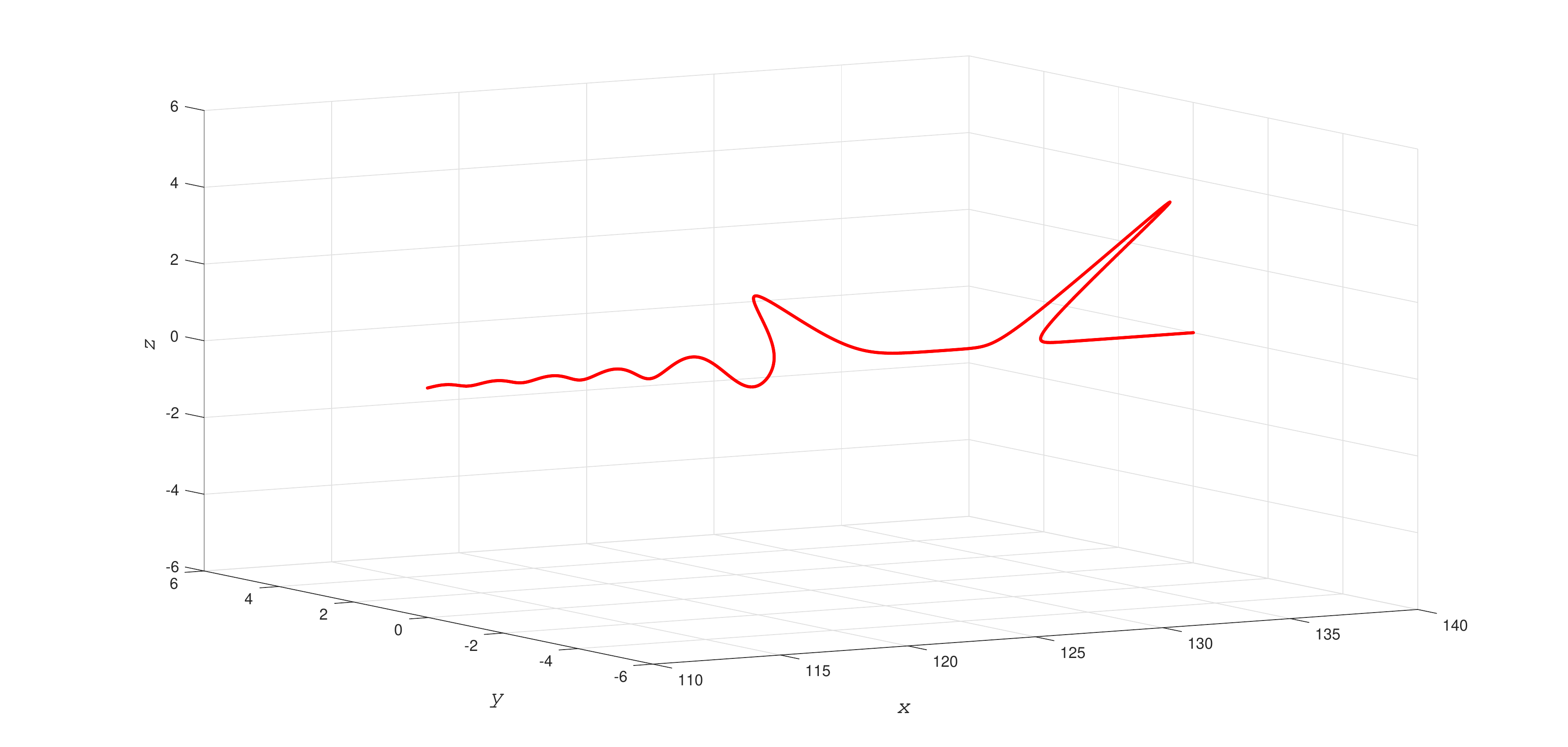}}}
	\caption{Snapshot at $t = 0.4$ representing the result of evolution of the helical soliton shown in Fig. \ref{f07}b) within the framework of non-integrable Eq.~(\ref{Eq01}). The initial helical soliton has desintegrated into several planar solitons propagating at different polarizations.}%
	\label{f09}
\end{figure}

When the initial condition was chosen in the form of helical soliton (\ref{Eq16}) with $\lambda = 2$ and $k = \lambda/\sqrt{3}$
(see Fig. \ref{f07}b), then in the integrable case of Eq.~(\ref{Eq02}), the soliton remains standing
in accordance with the theoretical prediction. However, when the same condition was used for the nonintegrable case of
Eq.~(\ref{Eq01}), it was observed that the helical soliton was disintegrated after a transient period
into several planar solitons propagating with the different angles to each other as shown in Fig. \ref{f09}.

The similar phenomenon was observed when the helical soliton (\ref{Eq16}) with $\lambda = 2$ and $k \approx 1.348$ (see Fig. \ref{f07}a) was substituted to the non-integrable Eq.~(\ref{Eq01}). Therefore, if the parameters of the initial condition even of a helical shape are far from the parameters of a stationary helical soliton of the non-integrable Eq.~(\ref{Eq01}), then instead of formation of a helical soliton the evolution results in the formation of a number of plane solitons propagating in different planes, so that the total helicity, including the helicity of a small-amplitude trailing wave, preserves.

\section{Discussion and conclusion}%
\label{Sect4}

We showed in this paper that helical solitons do exist within the framework of physically meaningful non-integrable
vector mKdV equation (\ref{Eq01}). The numerical approximations of such solitons have been developed
with the help of the dynamical system methods. As has been shown, the helical solitons exist as a result
of co-dimension one bifurcation along a curve in the parameter plane $(V,\omega)$. In particular,
such solitons can exist with only negative velocities in contrast to the traveling planar solitons
existing for positive velocities. Thus, the helical solitons appear to be similar to
breathers in the scalar mKdV equation \cite{Lamb1980}. Similar to the breathers in the mKdV equation,
we have shown that the helical solitons are stable in the time evolution of the vector mKdV equation (\ref{Eq01}).
This makes them interesting from the physical point of view.

The relevant solutions were compared to the helical solitons in the integrable vector
mKdV equation (\ref{Eq02}) which did not find applications in physical sciences.
In the latter case the helical solitons exists in a two-dimensional
region in the parameter plane $(V,\omega)$ and in particular, they can travel both with positive, zero or negative
velocities (i.e., being either ``subsonic'', ``sonic'' or ``supersonic''). When the helicity
is zero, the helical solitons reduce to the planar travelling solitons described by the scalar mKdV
equation  \cite{Lamb1980}.

Interaction between the travelling planar solitons has been investigated numerically in Ref. \cite{Nikitenkova2015}
within the framework of non-integrable vector mKdV equation (\ref{Eq01}) (see also \cite{Muslu2003, Uddin2013} and references therein). It was shown
that the interaction of travelling solitons is very nontrivial and inelastic, in
general. There is still an interesting open problem to investigate the interaction between the helical
solitons with the same and opposite helicity and between the helical and planar solitons. This problem will be considered elsewhere. 

In the conclusion, we mention that there several other vector-type equations (see, for example, \cite{Dmitriyev1993, Svin1994, Krylov1998, Yang2003, Tsuchida2015, Fench2017, Gromov2018}); some of them may possess helical soliton solutions. Among them there are equations of physical meaning \cite{Dmitriyev1993, Krylov1998, Yang2003, Gromov2018}, others are mainly of mathematical interest \cite{Svin1994, Tsuchida2015, Fench2017}.\\


{\bf Acknowledgement.} D.P. acknowledges the funding of this study from the State task program in the sphere of scientific activity of Ministry of Education and Science of the Russian Federation (Task No. 5.5176.2017/8.9). Y.S. acknowledges a support from the grant of President of Russian Federation for the leading scientific schools (NSH-2685.2018.5). \\

\end{document}